\documentstyle[epsfig,11pt]{article}
\def\be{\begin{eqnarray}}
\def\ee{\end{eqnarray}}

\def\half{{\textstyle \frac{1}{2}}}

\def\roughly#1{\mathrel{\raise.3ex\hbox{$#1$\kern-.75em%
\lower1ex\hbox{$\sim$}}}}

\def\bfpi{{\mbox{\boldmath $\pi$}}}
\def\bfalpha{{\mbox{\boldmath $\alpha$}}}
\def\bfSigma{{\mbox{\boldmath $\Sigma$}}}
\def\bfPi{{\mbox{\boldmath $\Pi$}}}
\def\bfGamma{{\mbox{\boldmath $\Gamma$}}}
\def\Tr{{\rm Tr}\,}

\begin{document}

\renewcommand{\thefootnote}{\arabic{footnote}}
\setcounter{footnote}{0}

\vskip 0.4cm
\hfill {\bf FZJ-IKP(TH)-1999-29}

\hfill {\today}
\vskip 1cm

\begin{center}
{\LARGE\bf Generalized Pions in Dense QCD}

\date{\today}

\vskip 1cm
Mannque Rho$^{a,b}$\footnote{E-mail: rho@spht.saclay.cea.fr},
Andreas Wirzba$^{c}$\footnote{E-mail: a.wirzba@fz-juelich.de}
and Ismail  Zahed$^{b}$\footnote{E-mail:
zahed@zahed.physics.sunysb.edu}

\end{center}

\vskip 0.5cm

\begin{center}

$^a$
{\it Service de Physique Th\'eorique, CE Saclay,
91191 Gif-sur-Yvette, France}

$^b$
{\it Department of Physics and Astronomy,
SUNY-Stony-Brook, NY 11794, U.\,S.\,A.}

$^c$
{\it FZ J{\"u}lich, Institut f\"ur Kernphysik (Theorie),
D-52425   J{\"u}lich, Germany}

\end{center}

\vskip 0.5cm

\begin{abstract}
QCD superconductors in the color-flavor-locked (CFL) phase
sustain light Goldstone modes (that will be referred to
as generalized pions) that can be
described as pairs of particle and/or hole excitations around
a gapped Fermi surface. In weak coupling and to leading
logarithm accuracy, their form factor,
mass and decay constant can be evaluated
exactly. These modes are found to satisfy an axial-Ward-identity,
constraining the mass of the Goldstone modes in the CFL phase.

\end{abstract}
\newpage

\renewcommand{\thefootnote}{\#\arabic{footnote}}
\setcounter{footnote}{0}

%%%%%%%%%%%%%%%%%%%%%%%%%%%%%%%%%%%%%%%%%%%%%%%%%%%%%%%%%%%%%%%%%%%%%%%%%%%
%%%%%%%%%%%%%%%%%%%%%%%%%% Introduction  %%%%%%%%%%%%%%%%%%%%%%%%%%%%%%%%%%
%%%%%%%%%%%%%%%%%%%%%%%%%%%%%%%%%%%%%%%%%%%%%%%%%%%%%%%%%%%%%%%%%%%%%%%%%%%

\centerline{\bf 1. Introduction}
\vskip 1cm

Quantum chromodynamics (QCD) at high density, relevant to the
physics of the early universe, compact stars and relativistic
heavy ion collisions, is presently attracting a renewed attention
from both nuclear and particle theorists. Following an early
suggestion by Bailin and Love~\cite{LOVE}, it was recently stressed
that at large quark density, diquarks could condense into a color
superconductor~\cite{ALL}, with novel phenomena.

At large density, quarks at the edge of the Fermi surface
interact weakly, although the high degeneracy of the Fermi
surface causes perturbation theory to fail. Particles can
pair and condense at the edge of the Fermi surface leading
to energy gaps. Particle-particle and hole-hole pairing
(BCS effect) have been extensively studied
recently~\cite{LOVE,ALL}. Particle-hole
pairing  at the opposite edges of the Fermi surface
(Overhauser effect)~\cite{OVERHAUSER} has also begun
to receive some attention~\cite{DGR92,SON2,US}. This is
however favored
only by a large number of colors~\cite{DGR92,SON2,US},
strong coupling (large gaps) or lower dimensions~\cite{US}.

The QCD superconductor breaks color and flavor symmetry spontaneously.
As a result, the ground state exhibits Goldstone modes that are either
particle-hole excitations (ordinary pions) or particle-particle and
hole-hole excitations (BCS pions) with a mass that vanishes in the
chiral limit. Effective-Lagrangian approaches to QCD in the color-flavor-locked
(CFL) phase have been discussed recently by some of us~\cite{Superqual} using
a nonlinear realization of spontaneously broken color-flavor symmetry, and
others~\cite{GATTO} using a linear realization with hidden gauge
symmetry. Both descriptions
are equivalent -- if vector dominance is exact --
due to the Stuckelberg mechanism~\cite{MEISSNER}. In general,
the effective Lagrangian approach provides a
convenient description of the long-wavelength physics based on
global flavor-color
symmetries, including flavor-color anomalies, but does not allow one
to determine the underlying parameters of the effective Lagrangian.
These parameters are important for a quantitative description of
the bulk (thermodynamic and transport)
properties of the QCD superconductor, including for instance the mass of
the recently discussed superqualiton~\cite{Superqual}. They
can only be determined using a more microscopic description of the
QCD superconductor.

In this letter, we will derive explicit expressions for the form
factor, temporal and
spatial decay constants and mass of the Goldstone modes in the weak coupling
regime in the CFL phase, and refer to~\cite{Superqual,GATTO} for the
discussion of the general aspects of the effective Lagrangian.
In section 2 we discuss the general features of the
QCD superconductor with screening. In section 3, we discuss
the bound state problem in the CFL phase, and derive explicit
results for the Goldstone modes.
In section 4, we derive a general axial-Ward-identity in the
QCD superconductor, constraining the  mass of the
Goldstone modes in weak coupling. Our conclusions are given in section~5.

\vskip 1.5cm
\centerline{\bf 2. QCD Superconductor}
\vskip 1cm

In the QCD superconductor, the quarks are gapped. Their propagation is
given in the Nambu-Gorkov formalism by the following matrix
\begin{eqnarray}
  {\bf S}=-i\langle{\bf \Psi}\,\overline{\bf\Psi}\rangle
  = \left(\begin{array}{cc} {S}_{11} & { S}_{12}\\
            {S}_{21} & {S}_{22}\end{array}\right)\;
 \label{PropMat}
\end{eqnarray}
in terms of the two-component Nambu-Gorkov
field ${\bf\Psi}=(\psi, \psi_C)$, where
$\psi$ refers to
quarks  and
$\psi_C(q)=C\bar \psi^T(-q)$ to charge
conjugated quarks,
respectively~\footnote{$q=(q^0,{\bf q})$ and $\bar\psi^T$ is the
transposed and conjugated field with $C\equiv i \gamma^2\gamma^0$.}.
According to Ref.~\cite{PisarskiSuperfluid}, the entries of ${\bf S}(q)$
in the massless case read~\footnote{We are
adopting the standard phase convention
between $\langle \psi
  \bar \psi\rangle$ and $S(q)$.}
\begin{equation}
\begin{array}{lclcl}
 S_{11}(q) &=& - i\langle\,\psi(q)\bar\psi(q)\,\rangle
           &=& \left[
                    \frac{\Lambda^{+}({\bf q})}{q_0^2-\epsilon_q^2}
                  + \frac{\Lambda^{-}({\bf q})}{q_0^2-{\bar\epsilon}_q^2}
                \right]\left(q\!\!\!/-\mu \gamma^0\right)   \; ,
 \\[1mm]
 S_{12}(q) &=& - i\langle\,\psi(q)\bar\psi_C(q)\,\rangle
            &=& - {\bf M}^\dagger\,  \left[
                 \frac{G^\ast(q)\,\Lambda^{+}({\bf q})}{q_0^2-\epsilon_q^2}
                  + \frac{{\overline G}^\ast(q)\,
                        \Lambda^{-}({\bf q})}{q_0^2-{\bar\epsilon}_q^2}
                \right]\; ,
 \\[1mm]
 S_{21}(q) &=& - i\langle\,\psi_C(q)\bar\psi(q)\,\rangle
           &=&   \left[
                    \frac{G(q)\,\Lambda^{-}({\bf q})}{q_0^2-\epsilon_q^2}
                 + \frac{{\overline G}(q)\,\Lambda^{+}
                   ({\bf q})}{q_0^2-{\bar\epsilon}_q^2}
                \right]\, {\bf M}\; ,
 \\[1mm]
 S_{22}(q)  &=& - i\langle\,\psi_C(q)\bar\psi_C(q)\,\rangle
            &=& \left(q\!\!\!/+\mu \gamma^0\right)\left[
 \frac{\Lambda^{+}({\bf q})}{q_0^2-\epsilon_q^2}
                  + \frac{\Lambda^{-}({\bf q})}{q_0^2-{\bar\epsilon}_q^2}
                \right] \; .
 \end{array}
  \label{PropFull}
\end{equation}
% $S_{22}=S_{11}^T$ and $S_{12}=S_{21}^T$.
% The quark
% propagator ${\bf S}$ has been discussed in~\cite{PISARSKI}.
Here
$\epsilon_q\equiv \mp\{\,(|{\bf q}|\mbox{$-$}\mu)^2
+{\bf M^\dagger}{\bf M} |G(q)|^2\,\}^{1/2}\approx  
\mp\{\,(|{\bf q}|\mbox{$-$}\mu)^2
+ |G(q)|^2\,\}^{1/2}$
are the energies  of a particle/hole~\footnote{This approximation
assumes  ${\bf M}^\dagger {\bf M} \approx {\bf 1}_{cf}$ in the
mass-shell condition.},
whereas the energies of an
antiparticle/hole are given by
${\bar\epsilon}_q\approx\mp \{\,(|{\bf q}|\mbox{+}\mu)^2
+|{\overline{G}}(q)|^2\,\}^{1/2}$~\cite{PisarskiSuperfluid,PISARSKI}.
The particle and antiparticle gaps are denoted by the complex-valued
functions  $G(q)$ and ${\overline{G}}(q)$, respectively.
The operators
$\Lambda^{\pm}({\bf q})=\half(1\pm \bfalpha\cdot\hat{{\bf q}})$ are the
particle/antiparticle projectors~\footnote{Note that
$\gamma^0\Lambda^{\pm}({\bf q})=\Lambda^{\mp}({\bf q})\gamma^0$,
$\gamma^5\Lambda^{\pm}({\bf q})=\Lambda^{\pm}({\bf q})\gamma^5$
and $\bfalpha\cdot \hat{\bf q} \,\Lambda^{\pm}({\bf q})= \pm |{\bf q}|\,
\Lambda^{\pm}({\bf q})$.}.  In the CFL phase
${\bf M}= \epsilon_f^a\epsilon_c^a\,\gamma_5$ with
$(\epsilon^a)^{bc}=\epsilon^{abc}$.
The charge conjugation operator $C$ is already
incorporated in the definition of the Nambu-Gorkov
field ${\bf\Psi}$.

\begin{figure}
 \centerline{\epsfig{file=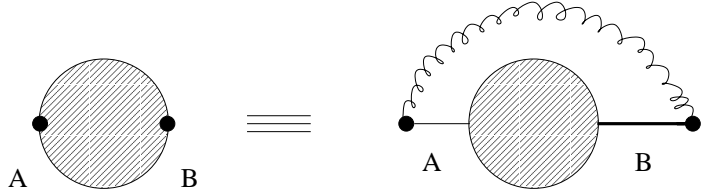,height=1in}}
 \caption{BCS gap equation. The thin and thick lines are the free
   and
dressed quark propagator, respectively, whereas the wiggly line
is the gluon propagator with A, B as Nambu-Gorkov indices.}\label{FigGap}
\end{figure}

For large
$\mu$, the antiparticles decouple:
$q_{||}\approx(|{\bf q}|-\mu)$ is the particle/hole momentum  at the
Fermi surface
in the direction of the Fermi momentum, such that
$\epsilon_q \approx \mp\sqrt{q_{||}^2+|G(q)|^2}$ and $\bar \epsilon_q
\approx \mp 2\mu$. Therefore,
we have
\begin{eqnarray}
  {\bf S}&\approx& \left(\begin{array}{rr}
     \gamma^0\,(q_0+q_{||}) \Lambda^{-}({\bf q})
    &    -{\bf M}^\dagger\,G^\ast(q)\Lambda^{+}({\bf q})
  \\
     {\bf M}\, G(q)\Lambda^{-}({\bf q})   & \gamma^0\,(q_0-q_{||})\
\Lambda^{+}({\bf q})
\end{array}\right)\,\,\frac{1}{q_0^2-\epsilon_q^2} \nonumber\\
%&=& \left(\begin{array}{cc}
%    \Lambda^{+}({\bf q}) \gamma^0\,(q_0+q_{||}) \Lambda^{-}({\bf q})
%    &    -\Lambda^{+}({\bf q}){\bf M}^\dagger\,G^\ast(q)
%          \Lambda^{+}({\bf q})
%  \\
%     \Lambda^{-}({\bf q}) {\bf M}\, G(q)   \Lambda^{-}({\bf q}) &
%\Lambda^{-}({\bf q})\gamma^0\,(q_0-q_{||})\Lambda^{+}({\bf q})
%\end{array}\right)\,\,\frac{1}{q_0^2-\epsilon_q^2} \; ,
\label{PropPart}
 \end{eqnarray}
with $q_0^2-\epsilon_q^2 \approx q_0^2 - q_{||}^2- |G(q)|^2$.
Using the color-identity
\begin{equation}
\sum_a \frac{\lambda^{aT}}2\,\epsilon^c \,\frac{\lambda^a}2
=-\frac 46\,\epsilon^c\,\,,
\label{identity}
\end{equation}
the gap function $G(q)$ in the CFL phase
satisfies (see Fig.~\ref{FigGap})
\begin{eqnarray}
 G(p) &=& \frac {4g^2}3 \,
 \int\frac{d^4q}{(2\pi)^4}\,i{\cal D}(p-q)\,
 \frac{G(q)}{q_0^2-\epsilon_q^2}\nonumber \\
     &=& \frac {4g^2}3 \,
 \int\frac{d^4q_{\mbox{\tiny $E$}}}{(2\pi)^4}\,{\cal D}(p-q)\,
 \frac{G(q)}{q_4^2+q_{||}^2+|G(q)|^2}\; .
 \label{GapEq}
\end{eqnarray}
The second expression refers to Euclidean coordinates. We note that
a similar equation is fulfilled by the antiparticle gaps through
$G(p)\rightarrow \overline{G}(p)$ on the left hand side of
(\ref{GapEq}) for the present approximations.
For perturbative screening,
the gluon-propagator in Euclidean space reads~\footnote{We are using a
simplified version as in~\cite{US}. To leading logarithm accuracy, the
results are unaffected.}
\begin{equation}
 {\cal D}(q) = \textstyle\frac12 \displaystyle \frac{1}{q^2 + m_E^2}
  + \textstyle\frac12 \displaystyle \frac{1}{q^2 + m_M^2} \; .
\label{GluonProp}
\end{equation}
Perturbative arguments give $m_E^2/(g\mu)^2=m_D^2/(g\mu)^2\approx
N_f/2\pi^2$ and $m_M^2/m_D^2\approx \pi{|q_4|/|4{\bf q}|}$, where
$m_D$ is the Debye mass, $m_M$ is the magnetic screening generated
by Landau damping and $N_f$ the number of flavors~\cite{LeBELLAC}.
Throughout we will refer to $m_M$ loosely as the magnetic screening
mass. We note that the perturbative screening vanishes at large $N_c$.
Nonperturbative arguments for screening~\cite{ZWAN} will not be
addressed here.

For a constant gap, (\ref{GapEq}) diverges logarithmically.
This is an ultraviolet effect that should not affect the
infrared behaviour at the Fermi surface~\cite{PISARSKI}.
With this in mind, we obtain
\begin{eqnarray}
  G(p_{||}) &\!\!\approx\!\!&
%  \frac{g^2\left(1+\frac{1}{N_c}\right)}{24 \pi^2}
  \frac{h_\ast^2}{6}
\int_0^{\infty}\! dq_{||}\,
   \frac{G(q_{||})}{\sqrt{q_{||}^2+|G(q_{||})|^2}}\nonumber\\
  &&\qquad\mbox{}\times
   \,\ln\left\{
   \left(1+\frac{\Lambda_{\perp}^2}
             {(p_{||}\mbox{$-$}q_{||})^2+m_E^2}\right)^3
   \left(1+\frac{\Lambda_{\perp}^3}{|p_{||}\mbox{$-$}q_{||}|^3
 +\frac{\pi}{4}m_D^2|p_{||}\mbox{$-$}q_{||}|}
    \right)^2\,\right\} \; ,\nonumber \\
 & &
 \label{GapEqScreened}
\end{eqnarray}
where
\begin{equation}
h_*^2=\frac 43 \frac {g^2}{8\pi^2}\,\, .
%\frac{{\rm min}(N_f,N_c)-1}{N_c} \; .
\label{HCFL}
\end{equation}
These relations can be readily generalized to arbitrary $N_c,N_f$ in the CFL
phase~\cite{US}.
The transverse cutoff $\Lambda_{\perp}=2\mu$ is exactly fixed in
weak coupling. Hence $\Lambda_{\perp}>m_E,m_M$ and the
logarithm in (\ref{GapEqScreened}) may not be expanded.
To leading logarithm accuracy, the gap equation (\ref{GapEq}) can be solved
using $x={\rm ln}(\Lambda_*/p)$ with
$\Lambda_*=(4\Lambda_{\perp}^6/\pi m_E^5)$ in the weak coupling
limit. The result is
\begin{equation}
 G(x)=G_0\,\sin (h_*\,x\,/\sqrt{3})
 \label{GapSolution}
\end{equation}
 with $G_0$ given by
\begin{eqnarray}
 G_0 \approx \left(\frac{4 \Lambda_{\perp}^6}{\pi m_E^5}\right)\,
  e^{-\frac{\sqrt{3}\pi}{2h_\ast}} \; .
\label{Gzero}
\end{eqnarray}
This result is the same as the one reached
in~\cite{SON1,PISARSKI,SchaferWilczek,US}.
We further note that the chiral condensate vanishes in the chiral limit,
\begin{equation}
 \langle\, \overline{\bf \Psi}\,{\bf \Psi}\,\rangle\,
 =i \int\frac{d^4q}{(2\pi)^4}\,\Tr {\bf S} (q) = 0\;
 \label{TraceS}
\end{equation}
because of the vector character of the interaction at the gap,
see e.g.\ (\ref{PropPart}).
However, the composite (quartic) chiral condensates do not, e.g.
\begin{equation}
 \left \langle\, \left(\overline{\bf \Psi}{\bf \Psi} \right)^2\,\right\rangle
 =\int\frac{d^4q}{(2\pi)^4}\,
 \Tr \left[{\bf S}(q)\, {\bf S}^\dagger(q) \right]\neq 0\,\,.
\label{TraceSS}
\end{equation}

\vskip 1.5cm
\centerline{\bf 3. Pions in the CFL Phase}
\vskip 1cm

\begin{figure}
 \centerline{\epsfig{file=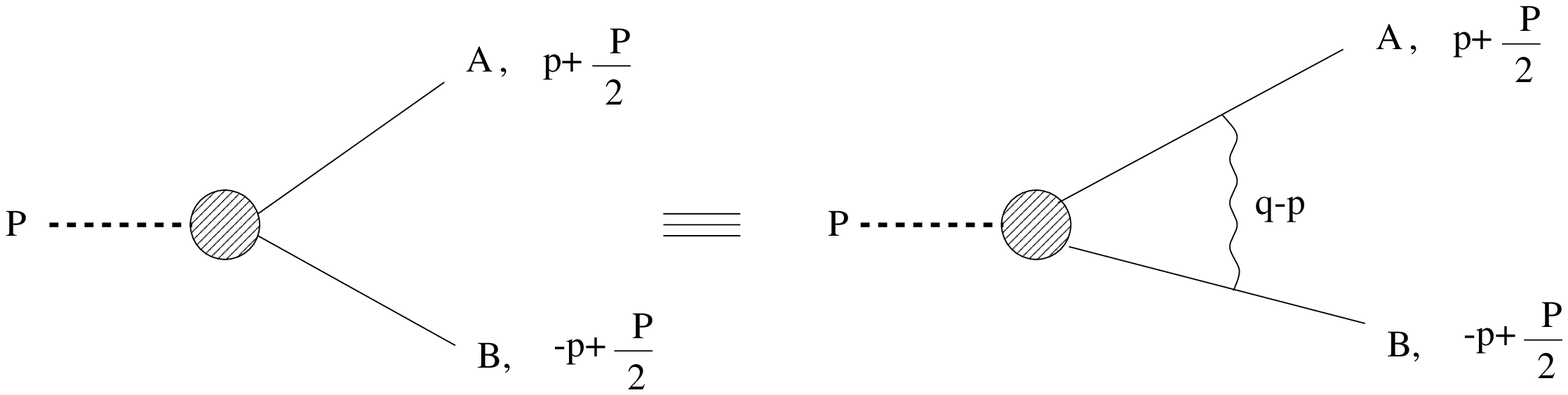,height=1.3in}}
 \caption{Bethe-Salpeter equation for the pions in the QCD
   superconductor.}
  \label{FigBetheSalp}
\end{figure}

At high density, QCD with $N_f=N_c=3$ and three degenerate
light quarks exhibits a phase with color-flavor locking that is
multiply degenerate~\cite{CFL}, i.e.
\begin{equation}
\langle\overline{\bf \Psi}\,
{\bf M}^{a\alpha}\left(e^{-i\gamma_5\pi^A
{\bf T}^A}\right)^{a\alpha}\,\rho_2\,{\bf \Psi}\,\rangle
\neq 0
\label{CFL1}
\end{equation}
with ${\bf M}^{a\alpha}=\epsilon^a_f\epsilon^{\alpha}_c\,\gamma_5
=({\bf  M}^\dagger)^{a\alpha}$,
$\rho_2$ a Pauli matrix active on the Nambu-Gorkov entries,
and ${\bf T}^A={\rm diag}\,(\tau^A,{\tau^{A}}^\ast)$ an $SU(3)_{c+F}$ valued
generator in the Nambu-Gorkov representation.
The CFL phase is invariant under the diagonal
of rigid vector-color plus vector-flavor, i.e. $SU(3)_{c+V}$.
The Goldstone modes in the CFL phase can be regarded as excitations with
particle and/or hole content. Their
wavefunction is driven by the Bethe-Salpeter kernel
shown in Fig.~\ref{FigBetheSalp}. Specifically,~\footnote{Note that
  the Fermion propagators are directed. Therefore, the two propagators
  incorporate  the total momentum
  $P$ with opposite sign and the relative momentum $q$ with the same
  sign.}
\begin{equation}
   \bfGamma^\alpha (p, P)=g^2\int\frac{d^4q}{(2\pi)^4}\,
 i{\cal D}(p-q)\,
i {\bf V}^a_\mu \,i {\bf S}(q\mbox{+}\frac P2)\,
\bfGamma^\alpha (q,P)\,
i{\bf S}(q\mbox{$-$}\frac P2)\,i {\bf V}_a^\mu
\label{t5}
\end{equation}
and the vertex is defined as
\begin{equation}
 {\bf V}^a_\mu \equiv \left(\begin{array}{cc} \gamma_\mu\lambda^a/2 & 0 \\
                 0 & C\left(\gamma_\mu\lambda^a/2\right)^TC^{-1}
                 \end{array} \right)
 = \left(\begin{array}{cc} \gamma_\mu\lambda^a/2 & 0 \\
                 0 &- \gamma_\mu{\lambda^{a\,T}}/2
                 \end{array} \right)
 \; .
\end{equation}
For $P=0$, the Bethe-Salpeter equation (\ref{t5}) admits the following
solution for the CFL pion vertex
\begin{eqnarray}
  {\bf \Gamma}^A \,(p,0)=
 i\gamma_5\,\frac {1}{F}\,\,
\left(\begin{array}{cc} 0 & \,- G^\ast(p)\,
      {{\bf M}^{A}}^\dagger \\
            G(p)\,{\bf M}^A & 0\end{array}\right)
\label{CFL2}
 \end{eqnarray}
with ${\bf M}^A={\bf M}^{a\alpha}\,( \tau^A)^{a\alpha}$ and
${{\bf M}^A}^\dagger={\bf M}^{a\alpha}\,( {\tau^A}^\ast)^{a\alpha}$.
Indeed, in terms of (\ref{CFL2}) the Bethe-Salpeter equation
for $P=0$ reduces to (\ref{t5}) with (\ref{HCFL}) for $h_*^2$,
after using
\begin{equation}
\sum_a \frac{\lambda^{aT}}2\,
\left( {\bf M}^A \right)\, \frac{\lambda^a}2 = -\frac 23\,{\bf
    M}^A\; , \quad
 \sum_a \frac{\lambda^{aT}}2\,
\left( {\bf M}\,{{\bf M}^{A}}^\dagger\,{\bf M}\right)\,
\frac{\lambda^a}2
\approx -\frac 23\,{\bf M}^A
\label{identity2}
\end{equation}
and ignoring the symmetric contributions in color-flavor which are
subleading in leading logarithm accuracy. Hence $G(p)=G^*(p)$
satisfies the gap equation~(\ref{GapEq}) and $F$ can be identified
as the pion decay constant. The distinction between temporal $F_T$
and spatial $F_S$ will be made below. We note that $G(p)$ in the
vertex (\ref{CFL2}) plays the role of the pion form factor in weak
coupling in the CFL phase. In the effective Lagrangian
approach~\cite{Superqual,GATTO} this feature is usually ignored by
treating the pions as point-like.

\begin{figure}
\centerline{\epsfig{file=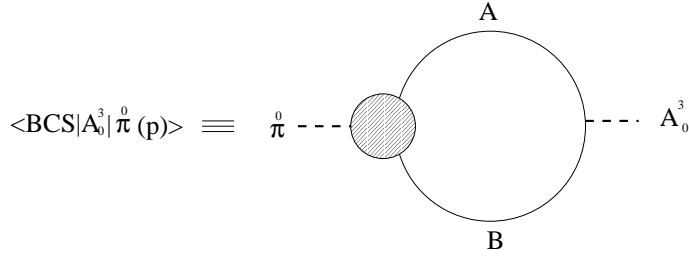,height=1.3in}}
\caption{Axial-vector transition in the QCD superconductor.}
\label{FigMEaxial}
\end{figure}

Having identified the Goldstone modes, we now proceed
to determine the pion decay constant $F$. For that we
use the standard definition in terms of the axial
vector current~\footnote{Note that because of the common term
  $+P_\mu$, the charged conjugated field in momentum space transforms
  as $\psi_C(q) = C\bar\psi^T(-q)$, although the corresponding field
  in position space transforms just as
$\psi_C(x) = C\bar\psi^T(+x)$; see Ref.~\cite{PisarskiSuperfluid}.}
\begin{equation}
\langle {\rm BCS} | {\bf A}_{\mu}^\alpha (0) |\pi^\beta_B (P)\rangle \equiv
iF\,P_{\mu}\,\delta^{\alpha \beta} \; .
\label{t7}
\end{equation}
In terms of the original quark fields $\psi$, the axial
vector current follows from Noether theorem. Hence
\begin{equation}
{\bf A}_{\mu}^{\alpha} (x) = \overline{\bf \Psi}
\gamma_{\mu}\gamma_5\frac 12\,{\bf T}^{\alpha}{\bf \Psi} \; .
\label{t8}
\end{equation}
In terms of (\ref{CFL2}) and (\ref{t8}) the relation (\ref{t7})
is given by the diagrams in Fig.~\ref{FigMEaxial}. Specifically,
\begin{equation}
iF\,P_{\mu}\delta^{\alpha\beta} =-\int \frac{d^4k}{(2\pi)^4}\,
\Tr\left(\gamma_{\mu}\gamma_5\,\frac 12\,{\bf T}^\alpha\,
i{\bf S}(k+\frac P2)\,i \bfGamma^\beta (k,P)\,i {\bf S}(k-\frac P2)\right)
\; .
\label{t9}
\end{equation}
In the (massive) pion rest frame $P=(M,0)$, (\ref{t9}) can be unwound for the
temporal component of the axial-vector current.
Expanding
the right-hand-side of (\ref{t9}) to leading order in $M$, yields
the temporal pion decay constant
\begin{equation}
 F_T^2\approx -8i\int\,\frac{d^4k}{(2\pi)^4}\,\frac{|G(k)|^2}
 {\left(k_0^2-\epsilon_k^2\right)^2}
 \; .
\label{t10}
\end{equation}
The result (\ref{t10}) is reminiscent of the result
for the Goldstone modes in the normal phase obtained with
the substitution $G(p)\rightarrow M(p)$ (constituent mass)~\cite{BERNARD}.
In the normal phase (with $\mu=0$), Lorentz symmetry is intact, so
$F_T=F_S$. At finite $\mu$, Lorentz symmetry is
upset~\cite{Kirchbach,Vesteinn,PisarskiTytgat}.
Indeed, the spatial component of (\ref{t9}) yields instead
\begin{equation}
 F_S^2\approx -8i\int\,\frac{d^4k}{(2\pi)^4}\,\frac{|G(k)|^2}
 {\left(k_0^2-\epsilon_k^2\right)^2}\,\,
 (\hat{\bf k}\cdot\hat{\bf P})^2 = \frac 13\, F_T^2 \; .
 \label{Fequx}
\end{equation}
In the CFL phase the Goldstone modes travel at a speed
less than the speed of light. The factor of $1/3$ in (\ref{Fequx})
is easily understood as the average of the current direction squared
over the Fermi surface.

For a constant gap, (\ref{t10}) diverges logarithmically. However,
this is an ultraviolet effect similar to the one already observed
in the gap equation~\cite{PISARSKI} that can be subtracted without
affecting $F$ at the Fermi surface at least to leading
logarithm accuracy. Assuming $G(k)\equiv G(k_{||})$, and performing
the integration over $k_0$ by contour with $\epsilon_k\rightarrow
\epsilon_k -i\epsilon$, we obtain
\begin{equation}
  F_T^2 \approx\frac {2\mu^2}{\pi^2}\int_0^\infty\,dk_{||} \,\,
    \frac{\,|G(k_{||})|^2}{\epsilon_k^3} \; .
  \label{t11}
\end{equation}
Inserting the leading logarithm solution (\ref{GapSolution})
to the screened gap equation (\ref{GapEqScreened})
in (\ref{t11})
and defining $x_0={\rm ln}(\Lambda _*/G_0)=\sqrt{3}\pi/(2h_*)$, we have
\begin{equation}
   F_T^2\approx \frac{2\mu^2}{\pi^2}
  \int_0^{x_0}\, dx\,e^{2\,(x-x_0)}\,\sin^2
   \left(\frac{\pi\,x}{2x_0}\right) =\frac{\mu^2}{2\pi^2}\,
\frac{8x_0^2+\pi^2\, (1-e^{-2x_0})}{4x_0^2+\pi^2}
\approx \frac {\mu^2}{\pi^2}\; .
\label{t12}
\end{equation}
Hence, $F_T^2/G^2_0\gg 1$, implying that the `size' of the pions
$r_{\pi}\approx 1/F_T$ in the CFL phase is very small. The inverse
size of the pion relates to the inverse transverse momentum exchanged
between pairs at the Fermi surface, which is of order $\mu$, 
irrespective of screening. Clearly $F_T$ vanishes if the BCS
gap vanishes through (\ref{t10}).

\vskip 1.5cm
\centerline{\bf 4. Axial Ward-identity}
\vskip 1cm

\begin{figure}
\centerline{\epsfig{file=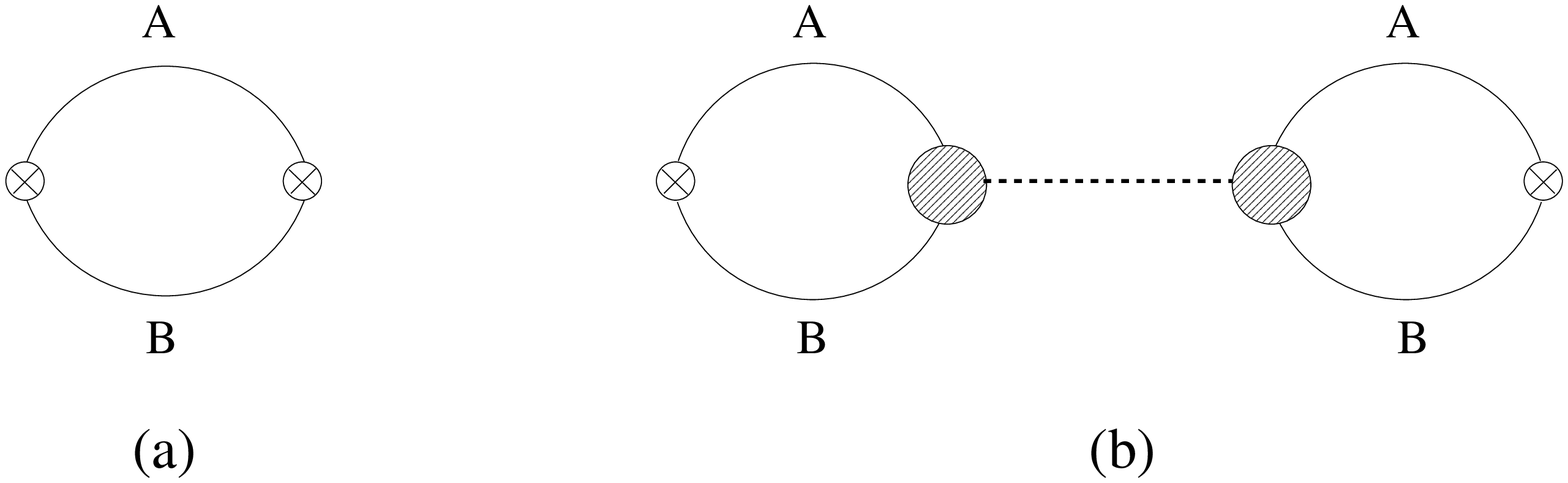,height=1in}}
\caption{(a) connected and (b) disconnected (generalized pion) contribution
in the QCD superconductor.}
\label{FigGOR}
\end{figure}

The underlying flavor symmetry of the QCD action entails
chiral Ward identities in the QCD
superconductor with relations between the mass and
decay constant of the Goldstone modes.
Indeed, when chiral symmetry is explicitly
broken by massive quarks $m_f=(m_u,m_d,m_s)$, then the pions
are expected to be massive. Hence
\begin{equation}
 0\equiv \int\, d^4x\,\partial^{\mu}_x\,
\left \langle {\rm BCS} \left| 
  T^*\,{\bf A}_{\mu}^\alpha (x) \,\bfpi_B^\beta (0)\,
                   \right|{\rm BCS}\right\rangle   \; ,
\end{equation}
where the axial-vector current ${\bf A}_{\mu}^a$
is given in (\ref{t8})
and the pion field ${\mbox{\boldmath $\pi$}}_B (x)$
in the QCD superconductor is defined as
\begin{eqnarray}
  \bfpi_B^\beta (x)= \left(\begin{array}{cc} 0 &
    \overline{\psi}\,\gamma^0\left(i\tau^\beta\gamma_5\,
 {\bf M}\right)^\dagger\gamma^0\,\psi_C (x) \\
           \overline{\psi}_C\,{\bf M}\, i\tau^\beta\gamma_5\,\psi (x) & 0
\end{array}\right)\,\,.
\label{a1}
\end{eqnarray}
The flavor axial-vector current in the CFL phase obeys
the local divergence equation
\begin{eqnarray}
\partial\cdot {\bf A}^\alpha (x) =
\left(\begin{array}{cc}
    \overline{\psi}\,i \half\, \left[m_f,\tau^\alpha\right]_{+}\,
 \gamma_5\,\psi (x) & 0\\
          0 & \overline{\psi}_C\, i\half\left[m_f
          ,{\tau^{\alpha}}^\ast\right]_{+}\, \gamma_5\,\psi_C (x)
\end{array}\right)\,\,.
\label{a2}
\end{eqnarray}
For massless quarks, the hermitean axial-isovector charge
\begin{eqnarray}
{\bf Q}_5^\alpha \equiv {\bf Q}_5^\alpha (x^0)= \int\, d^3x\,
\Tr\left(\begin{array}{cc}
    \overline{\psi}\half\tau^\alpha\gamma^0\gamma_5\,\psi (x) & 0 \\
         0&  \overline{\psi}_C\, \half{\tau^\alpha}^\ast\gamma^0\gamma_5\,\psi_C (x)
\end{array}\right)
\label{a3}
\end{eqnarray}
is conserved and generates axial-vector rotations, e.g.
\begin{equation}
   \left[{\bf Q}_5^\alpha , {\bf \Psi} (x)\right]
     =
   i\,\gamma_5\,\frac12\,{\bf T}^{\alpha} \,{\bf \Psi} (x)\,\,.
 \label{a4}
\end{equation}
In terms of (\ref{a2}-\ref{a4}), the identity (\ref{a1}) yields
the axial Ward-identity
\begin{equation}
 \int d^4x\,
\left 
 \langle {\rm BCS}\left|\,T^*\,
 \half\left[m_f ,\bfpi^\alpha(x)\right]_{+}\,
  \bfpi_B^\beta (0) \right|{\rm BCS}\right\rangle
 =\,  \left\langle {\rm BCS}\left| \bfSigma^{\alpha\beta}_B(0)
 \right|{\rm BCS}
 \right\rangle
  \; ,
 \label{a5}
\end{equation}
where the diquark field
$\bfSigma^{\alpha\beta}_B(x)$ is defined as
\begin{eqnarray}
  \bfSigma^{\alpha\beta}_B (x)= \left(\begin{array}{cc}
      0 &     \overline{\psi}\,\gamma^0
\left[\frac 12 \tau^\alpha,{\bf M}^\dagger\tau^\beta\right]_+\gamma^0\psi_C(x)\\
      \overline{\psi}_C
\left[\frac 12 \tau^\alpha, {\bf M}\,\tau^\beta\right]_+\psi (x) & 0\\
\end{array}\right)
\label{sigmaB}
\end{eqnarray}
and
$\bfpi (x)$ is the diagonal pion field
\begin{eqnarray}
  \bfpi^\alpha (x)= \left(\begin{array}{cc}
    \overline{\psi}i\tau^\alpha\gamma_5\,\psi (x) & 0\\ 0 &
          \overline{\psi}_C\,i{\tau^\alpha}^\ast\gamma_5\,\psi_C (x)
\end{array}\right)\,\,.
\label{a6}
\end{eqnarray}
The latter is to be contrasted with the off-diagonal or BCS pion field
(\ref{a1}). Clearly in the QCD superconductor,
$\bfpi (x)$ and $\bfpi_B (x)$ mix
through (\ref{a5}). This is expected, since particles and/or holes can
pop up from the superconducting state, thereby changing a normal pion
to a BCS pion. The true pion is a linear combination of both,
and the number of pseudoscalar Goldstone modes is only commensurate with
the dimension of the manifold spanned by (\ref{CFL1}).
A typical contribution to (\ref{a5}) is shown in
Fig.~\ref{FigGOR}a and \ref{FigGOR}b.
The dotted insertion in Fig.~\ref{FigGOR}b corresponds to the
BCS pion exchange in the superconductor.
 The nonconfining
character of the weak coupling description allows for the occurrence
of the gapped $qq$ and/or $\overline{q}q$ exchange of Fig.~\ref{FigGOR}a.
Hence,
\begin{eqnarray}
&\left\langle {\rm BCS}\left|\bfSigma^{\alpha\beta}_B(0) 
\right|{\rm BCS}\right\rangle
\approx -\int\!\! \frac{d^4q}{(2\pi)^4} \,{\rm Tr}\left[ i\gamma_5\,
\half\left[ m_f,{\bf T}^\alpha\right]_{+}\,
i{\bf S}(q)\,\bfPi_B^\beta\,
i{\bf S}(q)\right]\nonumber &\\
&-\left\{\int\!\! \frac{d^4q}{(2\pi)^4} \,{\rm Tr}\left[i\gamma_5
\half\left[m_f,{\bf T}^\alpha\right]_{+}
i{\bf S}(q)\,i\bfGamma^{\xi}_{\mbox{}}\,i{\bf S}(q)\right]
 \right\}\,\left(\frac{i}{M^2}\right)^{\xi\xi '}\,
\left\{\int\! \frac{d^4q}{(2\pi)^4} \,{\rm Tr}\left[i\bfGamma^{\xi '}\,
i{\bf S}(q)\,\bfPi_B^\beta\,
i{\bf S}(q)\right]\right\}\nonumber&\\
\label{a7}
\end{eqnarray}
with
\begin{eqnarray}
  \bfPi_B^\beta \equiv \left(\begin{array}{cc}
      0 &     \gamma^0\left(i \tau^\beta\gamma_5{\bf
      M}\right)^\dagger\gamma^0
  \\
     i \tau^\beta\gamma_5 {\bf M} & 0\\
\end{array}\right) \; .
\label{PiB}
\end{eqnarray}
In the chiral limit $m_i\rightarrow 0$, $i\in\{u,d,s\}$, 
the first term in (\ref{a7})
(Fig.~\ref{FigGOR}a) drops out and the identity is fulfilled if
$1/M^2$ is sufficiently singular in $m_i$ 
to match the numerator. The traces
can be evaluated in weak coupling. The result is~\footnote{The use of
$F_T$ instead of $F_S$ in the pion vertex follows from the fact that
the intermediate BCS pion is generated by a chiral rotation of the BCS
ground state. A 
similar interpretation in matter is made in~\cite{Vesteinn}.}
\begin{eqnarray}
&&\int \frac{d^4q}{(2\pi)^4}
\,\Tr\left[i\gamma_5\,\half\left[m_f,{\bf T}^\alpha\right]_{+}\,
i{\bf S}(q)\,i\bfGamma^{\xi}_{\mbox{}}\,i{\bf S}(q)\right]
={\cal O} (m_f^2)\; ,\nonumber\\
&&\int \frac{d^4q}{(2\pi)^4} \,\Tr\left[i\bfGamma^{\xi '}\,
i{\bf S}(q)\,\bfPi_B^\beta\,
i{\bf S}(q)\right]
=\delta^{\xi '\beta}\,\frac {16i}{F_T}\,
\int \frac{d^4q}{(2\pi)^4} \,
\frac{G(q)}{q_0^2-\epsilon_q^2}\; ,
\label{GOR10}
\end{eqnarray}
which shows that $M^2={\cal O}(m_f^2)$. To determine the coefficient,
we need to expand the vertices and the propagators in (\ref{a7}) to
leading order in $m_f$. The ${\cal O} (m_f)$ corrections to both $G(p)$
and $\Gamma (p)$ do not contribute. They trace 
to zero because of a poor
spin structure. Therefore, only the ${\cal O}(m_f)$ 
correction to the
propagator (\ref{PropFull}) is needed, i.e.
\begin{equation}
 \Delta {\bf S}(q) \approx 
%\frac{1}{2\mu\left(q_0^2 - \epsilon_q^2\right)}
    \left( \begin{array}{cc} 
     \frac{ m_f}{2\mu}\, \frac{q_0+q_{||}}{q_0^2 - \epsilon_q^2} & 
 \gamma^0 \left( \frac{m_f{\bf M}^\dagger \Lambda^+({\bf q})}{2\mu} 
   + \frac{{\bf M}^\dagger m_f \Lambda^-({\bf q})}{2\mu} 
\right)\,
  \frac{G^\ast(q)}{q_0^2-\epsilon_q^2}  \\
  \gamma^0 \left( \frac{m_f{\bf M} \Lambda^-({\bf q})}{2\mu} 
   + \frac{{\bf M} m_f\Lambda^+({\bf q})}{2\mu} \right)
  \frac{G(q)}{q_0^2-\epsilon_q^2} 
      &  \frac{m_f}{2\mu}\,\frac{-q_0 +q_{||}}{q_0^2 - \epsilon_q^2}
   \end{array}
 \right)\, .
\label{GOR11}
\end{equation}
Using (\ref{GOR11}) yields for the first trace in (\ref{GOR10})
\begin{eqnarray}
&& \int\,\frac{d^4q}{(2\pi)^4}
\,\Tr\left[i\gamma_5\,\half\left[m_f,{\bf T}^\alpha\right]_{+}\,
i{\bf S}(q)\,i\bfGamma^{\xi}_{\mbox{}}\,i{\bf S}(q)\right] \nonumber \\
&&=\ \frac{\mu G_0}{8 \pi^2 F_T}\,
{\rm Tr}_{cf}
\left( \left[ {m_f}^2,\tau^\alpha\right]
\left({\bf M}^\dagger {\bf M^\beta }- {{\bf M}^\beta}^\dagger{\bf M}
 \right) 
+ \left[ {m_f}^2,{\tau^\alpha}^\ast\right]
\left({\bf M}{\bf M^\beta}^\dagger- {{\bf M}^\beta}{\bf M}^\dagger
 \right) \right)\nonumber\\
\label{GOR12}
\end{eqnarray}
Using (\ref{GOR10}) and (\ref{GOR12}) in (\ref{a7}) and noting that
\begin{equation}
\left\langle \Sigma^{\alpha\beta}_B\right\rangle \equiv
{\rm Tr}\left(
\left[\frac {\tau^\alpha}2\,,
{\bf M}\,\tau^\beta\right]_+\,\rho_2\,{\bf S}\right)
=\delta^{\alpha\beta}\,8i\,\int\,
\frac{d^4q}{(2\pi)^4} \, \frac{G(q)}{q_0^2-\epsilon_q^2}
\label{GOR12x}
\end{equation}
we obtain for the mass of the Goldstone modes
\begin{eqnarray}
\left( M^2\right)^{\alpha\beta}&\approx&
\frac{\mu G_0}{4 \pi^2 F_T^2}
{\rm Tr}_{cf}
\left( \left[ {m_f}^2,\tau^\alpha\right]
\left({\bf M}^\dagger {\bf M^\beta }
 \mbox{$-$}{{\bf M}^\beta}^\dagger{\bf M}
 \right) 
+ \left[ {m_f}^2,{\tau^\alpha}^\ast\right]
\left({\bf M}{\bf M^\beta}^\dagger
 \mbox{$-$}{{\bf M}^\beta}{\bf M}^\dagger
 \right) \right)\nonumber\\
&\approx& \sqrt{\frac{2}{3}}\,\frac{256 \pi^4 }{9  g^5}\, 
\exp\left(-\frac{3\pi^2}{\sqrt{2}\,g} \right)\,
\left\{
{\rm Tr}_{cf}
\left( \left[ m_f^2 ,\tau^\alpha\right]
\left({\bf M}^\dagger {\bf M^\beta }- {{\bf M}^\beta}^\dagger{\bf M}
 \right) \right) \right.  \nonumber \\
&&\qquad   \left. \mbox{}+
 {\rm Tr}_{cf}\left( \left[ m_f^2, {\tau^\alpha}^\ast
 \right]
\left({\bf M}{\bf M^\beta}^\dagger- {{\bf M}^\beta}{\bf M}^\dagger
 \right) \right)\right\}
\label{GOR13} 
\end{eqnarray}
where (\ref{t12}) and   (\ref{Gzero}) 
were inserted into the last line for $F_T^2$ and $G_0$, respectively.
Furthermore, 
$m_E= \sqrt{(3/2)}\, g\mu/\pi$ was used.
The $m_f^2$ behaviour is consistent with the one 
suggested in~\cite{Pisarski:1999cn}. The color-flavor trace
in (\ref{GOR13}) vanishes, suggesting that the generalized pions remain
massless to order $m_f^2$ in the CFL phase
~\footnote{This result is exact in mass perturbation theory in the CFL phase
modulo footnote 3.}.

\vskip 1.5cm
\centerline{\bf 5. Conclusions}
\vskip 1cm

We have discussed certain bulk features of the QCD superconductor.
In the CFL phase, the order parameter is multidegenerate leading
to Goldstone modes, with temporal and spatial decay constants
that can be calculated exactly in weak coupling. We find
$F_T^2/G_0^2\gg 1$ and
$F_S^2/F_T^2=1/3$. The Goldstone modes  have a very small size
and propagate with a speed that is less than the speed of light.
The multidegeneracy of the Goldstone manifold is lifted by finite quark masses.
The Goldstone modes are found to obey a generalized axial Ward identity,
constraining the mass of the pion in the CFL phase. 
We note that the small size of the pion implies
that the recently discussed superqualitons~\cite{Superqual}
are in general heavy, with $M_s/G_0\approx (F_S/G_0)^2\gg 1$.
The mismatch between the temporal and spatial
decay constants may be relevant for soft pion emission in cold and dense matter.
Further issues regarding the CFL spectrum will be discussed elsewhere.

\vskip 1.5cm
\section*{Acknowledgments}
\vskip .5cm
IZ thanks Edward Shuryak for  discussions. We thank Y. Kim for help
with the Figures. This work was supported in part by US-DOE DE-FG-88ER40388
and DE-FG02-86ER40251.

\vskip 1.5cm
{\it Note Added:\,\, After completion of our work Ref.~\cite{SS} appeared
where similar issues were addressed using different arguments.}

\end{document}